\documentclass[conference]{IEEEtran}
\pdfoutput=1

\usepackage[T1]{fontenc}
\usepackage{amssymb}	
\usepackage[pdftex]{graphicx}	
\usepackage{hyperref,xcolor}

\usepackage{mathtools}
\usepackage[lined,ruled,linesnumbered]{algorithm2e}

\usepackage{fancyhdr}
\newcommand{\bb}{\boldsymbol{b}}
\newcommand{\bv}{\boldsymbol{v}}
\newcommand{\by}{\boldsymbol{y}}
\newcommand{\bx}{\boldsymbol{x}}
\newcommand{\bn}{\boldsymbol{n}}
\newcommand{\bH}{\boldsymbol{H}}

\newcommand{\bT}{\boldsymbol{T}}
\newcommand{\bi}{\boldsymbol{i}}
\newcommand{\be}{\boldsymbol{e}}

\newcommand{\bu}{\boldsymbol{u}}

\newcommand{\bX}{\boldsymbol{X}}

\newcommand{\bW}{\boldsymbol{W}}

\newcommand{\bV}{\boldsymbol{V}}

\newcommand{\bU}{\boldsymbol{U}}
\newcommand{\bt}{\boldsymbol{t}}
\newcommand{\bs}{\boldsymbol{s}}
\newcommand{\bphi}{\boldsymbol{\phi}}

\usepackage{amsmath}
\usepackage{algorithmic}

\begin{document}

\title{\vspace{-1cm}\textsf{\small Submitted to EUSIPCO 2016}\bigskip \\
Distributed multi-frequency image reconstruction for radio-interferometry}
\author{
\IEEEauthorblockN{Jérémy Deguignet, André Ferrari, David Mary and Chiara Ferrari} 
\IEEEauthorblockA{Lab. J.-L. Lagrange, Université de Nice Sophia Antipolis, CNRS, Observatoire de la Côte d'Azur, \\ Parc Valrose, F-06108 Nice cedex 02, France}
}

\maketitle

\begin{abstract}
The advent of enhanced technologies in radio interferometry and the perspective of the SKA telescope 
bring new challenges in image reconstruction. One of these challenges is the spatio-spectral reconstruction of large (Terabytes) data cubes with high fidelity.
This contribution proposes an alternative implementation of one such 3D prototype algorithm, MUFFIN (MUlti-Frequency image reconstruction For radio INterferometry), which combines spatial and spectral analysis priors. Using a recently proposed primal dual algorithm, this new version of MUFFIN  
allows a parallel implementation where  computationally intensive steps are  split by spectral channels. 
This parallelization allows to implement computationally demanding translation invariant wavelet transforms (IUWT), as opposed to the union of bases used previously.
This alternative implementation is important as it opens the possibility of  comparing  these efficient  dictionaries, and others, in spatio-spectral reconstruction. 
Numerical results show that the IUWT-based version can be successfully implemented at large scale with performances comparable to union of bases.
\end{abstract}
\IEEEpeerreviewmaketitle

\section{Introduction}
Imaging reconstruction algorithms for radio-interferometry have experienced an important growth over the past decade.
This research activity is constantly stimulated by methodological advances in inverse problems and optimization on the one hand,
and by recent technological advances in phased arrays on the other hand. In the framework of an international, extremely ambitious
 large scale radio phased array to  be built in the next years, the Square Kilometer Array  (SKA, see \cite{Dewdney13}), an 
 increasing number of researchers in signal processing and radio astronomy join in common efforts. With thousands of dishes and
 millions of dipoles spread over hundreds of kilometers, the SKA poses a number of high-level challenges to several research domains. 
 One of these challenges is the ability to reconstruct high-fidelity spatio-spectral  data cubes (multifrequency images) of several TeraBytes (TB).

In radio interferometry, the receivers can be classical dishes or groups of co-phased sensors (dipoles) called stations. The spatial position 
of a pair of receivers defines one of the baselines of the telescope array. In the ideal case, two receivers with baseline $\bb$ observing in a narrow frequency band   $\nu=c/\lambda$ measure a complex visibility, $v_{\lambda}$, which corresponds to a sample of the Fourier spectrum of the intensity distribution of interest at spatial frequency $\bb/\lambda$. The sampling of the Fourier space is thus governed by
the configuration of the receivers in the radio interferometer geometry.  Successive snapshots measurements
increase the coverage of the Fourier space because the Earth rotation modifies the  configuration of the array baselines with respect to the sky.

The SKA is emblematic of a new generation of low frequency radio telescopes, which are able to
provide unprecedented sensitivity, resolution and large fields of view (as already demonstrated, for instance, by the SKA pathfinder
LOFAR (Low Frequency Array) \cite{Lofarcourt}).  
Ultimately, the SKA will achieve a tremendously broad Fourier frequencies coverage allowing (sub-)arcsec  resolution over hundreds of frequency bands and a dynamic range expected to cover up to seven orders of magnitude (see Table 1 in \cite{Dewdney13}).
But this evolution has a price: image reconstruction algorithms must be able to 
process  in a manageable amount of time huge amounts of data (leading e.g. to storage and memory issues), especially in a multifrequency framework. 
Indeed, performing a joint reconstruction of both spatial and spectral behaviors of  radio sources is a key issue  
to fully characterize such sources  \cite{krausb}.  The spatio-spectral models used to achieve high fidelity reconstruction is also a key issue. The goal
of the present work is to allow large scale comparison between two state-of-the-art  approaches, both based on sparse priors but expressed through different types of redundant dictionaries (Isotropic Undecimated Wavelet Transforms, IUWT \cite{Starck07},  and union of bases).

So far, however,  existing image reconstruction algorithms are mostly monochromatic.
Interestingly, sparsity was early recognized  as a powerful principle for reconstruction and has lead to the most populated family of imaging algorithms. 
Their patriarch is the CLEAN algorithm (\cite{clean}, devised in 1974), which expresses and exploits the  sparsity of the sky intensity distribution  in the canonical basis. 
Efficient monochromatic algorithms relying on more general sparse models (through redundant dictionaries) have since then proven their efficiency in radio imaging:
recent examples include the  works \cite{Garsden15} (IUWT), \cite{SARA,Purify,Onose16} (union of bases), which rely on  global minimization of sparsity-regularized functionals, or \cite{Moresane} (IUWT), which combines complementary types of sparse recovery methods in a greedy manner. 

Turning  to the few existing multi-frequency reconstruction algorithms,
most of the proposed approaches rely on a physical model for the frequency-dependent brightness distribution. In  \cite{RauMSMF}, a Taylor expansion of a power-law is adopted to model the  flux dependence in frequency of astrophysical radio sources.
More recently,  reconstruction algorithms relying on parametric models for this dependence have been proposed.
In  \cite{Junklewitz14},   the authors propose to   address the estimation problem using a Bayesian framework.   
The works \cite{Bajkova11} propose a   constrained maximum entropy estimation algorithm in order to account for the frequency dependence of the intensities.

These ``semi-parametric'' methods rely on spectral models and thus clearly offer advantages and estimation accuracy when the model is indeed appropriate. However, across the broad frequency coverage of current radio facilities, radio sources exhibiting complex spectral shapes (not simple power laws) are expected. For instance, the works  \cite{Kellerman74} evidence that some sources may exhibit one or more relative minima, breaks and turnovers. For the new generation of low frequency telescopes  such as LOFAR, recent studies have also shown that second order broadband spectral models are often insufficient  \cite{Scaife12}.  Attempts  to relax the spectral power-law model are thus necessary. One  such attempt, in \cite{Wenger:2014ty}, formulates the problem as an inverse problem with a smooth spectral  regularization allowing for local deviations. The present study is another such attempt.

In \cite{muffinv1}, the authors proposed to reconstruct a multi-wavelength sky image 
using a fully non-parametric approach. The resulting algorithm (named MUFFIN for MUlti-Frequency image reconstruction 
For radio INterferometry) performs a joint spatio-spectral multi-wavelength reconstruction by
 incorporating a spectral regularization. As mentioned above,  the spectral dimension
critically blows up the size of the inverse problem, with targeted sizes reaching 80 TB for SKA cubes.
To cope with computational issues, optimization in MUFFIN  was implemented using the  alternative direction method of multipliers (ADMM)
\cite{muffinv1}. However, identified limitations of MUFFIN are {(i) the resolution of a large size linear system at each iteration
(ii) the high number of primal and dual variables, generating   expansive   memory costs.}

In the present work,  a new implementation of MUFFIN is derived using the primal-dual algorithm proposed in \cite{Condat,Vu11}.
This new implementation presents three main advantages: \\ { (i) it uses a reduced number of variables (lower memory costs); (ii) it avoids the resolution of linear systems;\\
(iii) it 
allows a more efficient parallelization (computationally intensive steps are parallelized by wavelengths).} \\These processing improvements allow 
to implement computationally demanding non-orthogonal wavelet transforms and associated exact adjoint operators (IUWT). This makes future studies in position
of comparing approaches based on IUWT vs union of bases in a large scale 3D framework. 

The paper is organized as follows: section~\ref{section2} introduces the model and the inverse problem. The optimization algorithm is presented in section~\ref{section4}
and its performances are illustrated in section~\ref{section5}. 

\vspace{-4mm}
\section{Spatio-spectral inverse problem for radio-interferometry}
\label{section2}
\vspace{-4mm}
A complex visibility measures the spatial coherence of the electric field at the position of two antennas
and wavelength $\lambda$. 
Noting  $w$ the coordinate along the line of sight and  $(u,v)$ the coordinates in its perpendicular plane, 
the visibility $v_\lambda (u,v,w)$ is related to the sky brightness distribution at $\lambda$, $x^\star_\lambda(l,m)$, by:
\begin{equation}
v_\lambda (u,v,w) = \int \int \frac{x^\star_\lambda(l,m)}{\sqrt{1 - l^{2} - m^{2}}} e^{-\frac{2\pi}{\lambda} i(ul+vm+wn)} \textrm{d}l \textrm{d}m
\label{viseq}
\end{equation}
with $l^{2} + m^{2}+ n^{2} = 1$. 
When the term ${w\sqrt{1-l^2-m^2}}$ can be considered small (for instance for coplanar baselines or very small fields of view), (\ref{viseq}) reduces to a Fourier transform.
In the general case, this term induces a form of
 non-isoplanatism, as it introduces
a   direction dependent effect (DDE). 
In practice, other DDE  (e.g. ionosphere or antenna/station beam) exist. They are assumed to be calibrated in this study.
 
Combining  all the measured complex visibilities (resp. the discretized sky brightness image) in a vector $\bv$ (resp. $\bx{\bf{^\star}}$) 
and omitting for now the dependence in $\lambda$,  model (\ref{viseq}) can be expressed as:
\begin{equation}
\bv = \bphi \bx{\bf{^\star}} + \be
\end{equation} 
where $\bphi$ is a linear mapping from the image domain to the visibilities, which includes the DDE and $\be$ is
a noise vector.
Various iterative algorithms such as  \cite{wprojection,wsclean} have been proposed in the literature
to compensate for these DDE. They lead to the image plane model:
\begin{equation}
\label{eq:pb}
\by = \bH \bx{\bf{^\star}} + \bn
\end{equation}
where $\by$ is the so-called dirty image and $\bH$ is a convolution operator. This model assumes that DDE have been corrected for, or
that they lead to point spread functions (PSF) that are piecewise constant across the field.
From now on,  model (\ref{eq:pb}) will be considered.

Let $\bx_{l}{\bf{^\star}}$ be the column vector collecting the sky intensity image at 
wavelength $\lambda_{l}$, with $l=1,\ldots,L$ and $L$  the  number of spectral channels. The dirty image 
$\by_{l}$ at 
wavelength $\lambda_{l}$ is related to the sky intensity image by:
\begin{equation}
\label{model}
    \by_{l} = \bH_{l} \bx_{l}{\bf{^\star}} + \bn_{l}
\end{equation}
where $\bn_{l}$ is a perturbation vector accounting for noise and model error   and $\bH_{l}$ represents  convolution by the PSF at $\lambda_{l}$.

Eq.~(\ref{model}) defines a linear inverse problem, which is ill-posed owing to the partial coverage of the Fourier plane. {
This problem can be solved in a cost minimization framework, by adding to the data fidelity term a regularization term $f_\text{reg}$ related to some prior on $\bx_1{\bf{^\star}},\ldots,\bx_{L}{\bf{^\star}}$. Let  $\bx_1,\ldots,\bx_{L}$ denote the corresponding optimization variables at each wavelength
and $\bX$ denote the concatenation matrix $ \bX := [\bx_1,\ldots,\bx_{L}]$.
 With these notations the  cost function  writes:
\begin{equation} 
\label{eq:reg}
\min_{\bX}\; \sum_{l=1}^{L} \frac{1}{2 \sigma_{l}^{2}}  \|\by_{l} - \bH_{l} \bx_{l} \|^2 + f_\text{reg}(\bX)
\end{equation}
where $\sigma_{l}^2$ is the noise variance of the corresponding dirty image $\by_{l}$.
}
Note that the fidelity term in (\ref{eq:reg}) separates in wavelengths. This is justified if the width of the spectral PSF is 
sufficiently small so that fluxes in adjacent spectral channels are not mixed up.

Several recent works have shown that regularization based on  sparse 
representations in  appropriate transform domains 
can be very effective. Such regularization terms can be formulated in an analysis or in a synthesis
framework.  These two formalisms are discussed and compared e.g. in \cite{elad2007}. 
For both approaches, redundant dictionaries improve over non redundant (orthogonal) ones.
For narrow-band radio-interferometric imaging, state-of-the-art results appear so far to be obtained with union of bases \cite{SARA,Purify,Onose16} and IUWT  \cite{Garsden15,Moresane}.

In complement to the  classical positivity constraint $\boldsymbol{1}_{\mathbb{R}^+}(\bX)$, the present study opts for a sparse analysis prior operating both spatially and spectrally, leading to a regularization of the form:
\begin{equation}
f_\textrm{reg}(\bX) := \boldsymbol{1}_{\mathbb{R}^+}(\bX) + 
\mu_{s}  \sum_{l=1}^{L}   \| \textbf{W}_s \bx_{l} \|_1 +
\mu_{\lambda} \sum_{n=1}^{N} \| \textbf{W}_\lambda \bx^{n} \|_1	 
  \label{eq:RegAll}
\end{equation}
In (\ref{eq:RegAll}),
  $\bx_l$  (the $l^{th}$ column of $\bX$) corresponds to the image at  wavelength $l$ and $\bx^n$ (the $n^{th}$ row of $\bX$) is the spectrum associated to  pixel $n$. 
$\textbf{W}_s$ and $\textbf{W}_\lambda$ are the operators associated  with, respectively, the spatial and spectral 
decomposition. IUWT 
will here be considered for the spatial regularization and  a cosine decomposition for the spectral model.
It is also important to underline the central role of the last regularization term with parameter $\mu_\lambda$ in (\ref{eq:RegAll}). This term prevents the  
optimization problem (\ref{eq:reg}) from being separable w.r.t. the $\bx_{l}$. This makes the sparse spatial and spectral priors linked together and the regularization truly spatio-spectral. 

As far as large scale implementation is concerned, another important point is that in (\ref{eq:RegAll}) the first and second terms are separable w.r.t. the wavelengths while the last term is separable w.r.t. the pixels. Note finally that, similarly to \cite{lsf}, a synthesis approach could have been considered for the spectral regularization. In this case, however, the fidelity term in (\ref{eq:reg}) would be no more separable w.r.t. the wavelengths.
\vspace{-2mm}
\section{Optimization algorithm}
\label{section4}

The works \cite{muffinv1} proposed to minimise the convex  problem described by (\ref{eq:reg}, \ref{eq:RegAll}), 
using an ADMM algorithm. {A major drawback of this solution is the necessity to solve a large linear 
system at each iteration. 
This was kept computationally tractable in \cite{muffinv1,Purify} by using for $\bW_s$ a concatenation
of orthogonal wavelet bases.
Another drawback is }the amount of memory required by the multiplication of the primal and dual variables, which are each  of the size of the data cube (at least; this reaches several data cubes for redundant analysis coefficients). 
In order to reduce the required memory, this communication proposes to replace the ADMM algorithm by
the primal-dual optimization algorithm \cite{Condat,Vu11}. It proceeds by full splitting of the inverse problem and so  can call individually each proximal operator of the functions.
Application of \cite{Condat,Vu11} to (\ref{eq:reg}, \ref{eq:RegAll})  leads to Algorithm \ref{algo1}, where: 
\begin{alignat}{2}
\text{sat}(u) := 
\begin{dcases}
  -1 & \quad \text{if} \quad u < -1\\
  1 & \quad \text{if} \quad u > 1\\ 
  u &  \quad \text{if} \quad |u| \leq 1
\end{dcases}
\end{alignat}
and $(\cdot)_+$  is the projection on the positive orthant.
Parameters $\rho$, $\sigma$ and $\tau$ are fixed according to \cite{Condat}
in order to guarantee the convergence of the algorithm.
 
\begin{algorithm}
\SetKwInOut{Input}{Initialize}
\SetKwInOut{Output}{Return}
\Input{$\bx$, $\textbf{U}$ and $\textbf{V}$}
\Repeat{stopping criterion is satisfied.}{
  $\boldsymbol{\nabla} = \left(\bH_1^\dagger(\bH_1\bx_1-\bi_1^d)  \,|\, \cdots \,|\, \bH_L^\dagger(\bH_L\bx_L-\bi_L^d)\right)$\;
  $\tilde{\bX} = \left(\bX - \tau(\boldsymbol{\nabla}+\mu_s\bW_s^\dagger \bU + \mu_\lambda\bV\bW_\lambda^\dagger)\right)_+$ \;
   $\tilde{\bU}=\text{sat}\left(\bU + \sigma\mu_s\bW_s(2\tilde{\bX} -\bX)\right)$\;
   $\tilde{\bV}=\text{sat}\left(\bV + \sigma\mu_\lambda\b(2\tilde{\bX} -\bX)\bW_\lambda\right)$\;
  $(\bX,\bU,\bV) = \rho (\tilde{\bX},\tilde{\bU},\tilde{\bV})+ (1-\rho)(\bX,\bU,\bV)$\;
}
\Output{$\bX$}
 \caption{MUFFIN algorithm. \label{algo1}}
\end{algorithm}
Note that Algorithm \ref{algo1} requires 6 variables  ($\tilde{\bX}, \bX, \tilde{\bU}, \bU, \tilde{\bV}, \bV$) in addition to the gradient
and 5 if $\rho = 1$, while 10 variables are necessary in \cite{muffinv1}.
Moreover, in contrast to \cite{muffinv1,Purify}, Algorithm 1 does not require to solve at each iteration
large linear systems.  This allows  the use of highly redundant, 
translation invariant wavelet transforms  like (for instance) IUWT \cite{Starck07}.

A major advantage of Algorithm \ref{algo1} is that the most demanding steps are separable w.r.t.
the wavelengths, leading to the following parallel implementation.
MUFFIN is distributed on a cluster where the master node centralises the reconstructed data cube and
each wavelength is associated to a compute node $l$. The algorithm iterates as follows:
\begin{enumerate}
\item The master node computes $\bT = \mu_\lambda\bV\bW_\lambda^\dagger$ and sends the column $l$ of $\bT$,
denoted as $\bt_l$, to node $l$.

\item Each node $l=1\ldots L$ computes sequentially:
\begin{align}
&\boldsymbol{\nabla}_l=\bH_l^\dagger(\bH_l\bx_l-\by_l) \label{grad} \\
& \bs_l=\mu_s\bW_s^\dagger \bu_l \label{adjointspat}\\
&\tilde{\bx}_l = \left(\bx_l - \tau(\boldsymbol{\nabla}_l+\bs_l + \bt_l)\right)_+ \\
&\tilde{\bu}_l=\text{sat}\left(\bu_l + \sigma\mu_s\bW_s(2\tilde{\bx}_l -\bx_l)\right) \label{anaylsespat}\\
&(\bx_l,\bu_l) = \rho (\tilde{\bx}_l,\tilde{\bu}_l)+ (1-\rho)(\bx_l,\bu_l)  \label{updatenode}
\end{align}

\item Each node sends $\bx_l$ and $\tilde{\bx}_l$ to the master and the master computes sequentially:
\begin{align}
& \tilde{\bV}=\text{sat}\left(\bV + \sigma\mu_\lambda(2\tilde{\bX} -\bX)\bW_\lambda\right) \\
& \bV = \rho \tilde{\bV}+ (1-\rho)\bV
\end{align}
\end{enumerate}
Note that the particularly time consuming steps associated to (\ref{grad},\ref{adjointspat},\ref{anaylsespat}) are computed in parallel at each wavelength.
This is particularly important for (\ref{adjointspat}) when the transform is not orthogonal. In such cases,
the adjoint operator differs from the perfect reconstruction synthesis operator and its implementation may not benefit from the same fast algorithm.

A distributed memory implementation of MUFFIN will be available\footnote{\texttt{https://github.com/andferrari/muffin.jl}}.
The algorithm has been implemented in Julia \cite{Julia}, which provides a multiprocessing environment based on message passing. 

\section{Simulations}
\label{section5}

\begin{figure}
\centerline{
	{\includegraphics[width=.45\columnwidth]{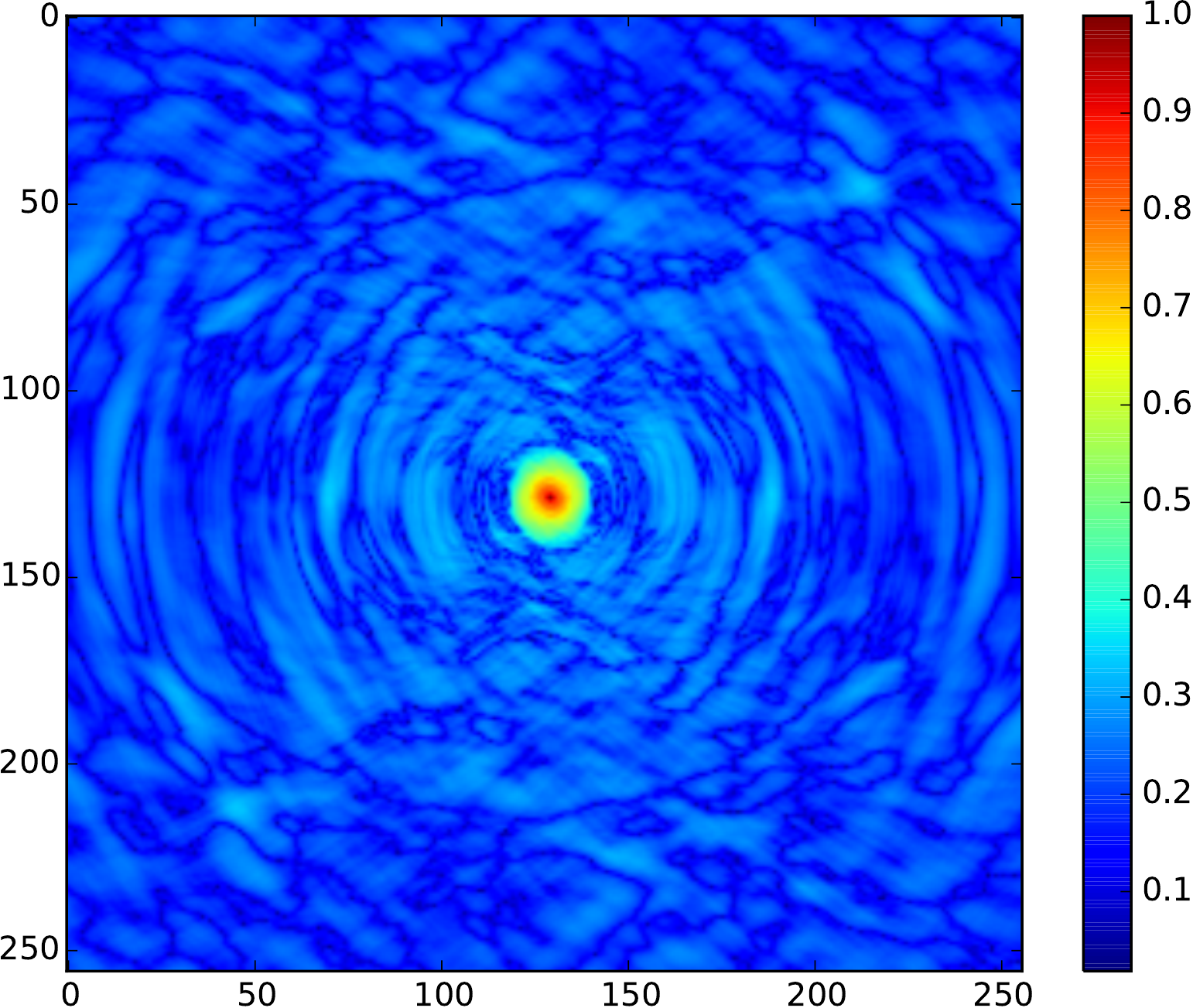}}
		{\includegraphics[width=.45\columnwidth]{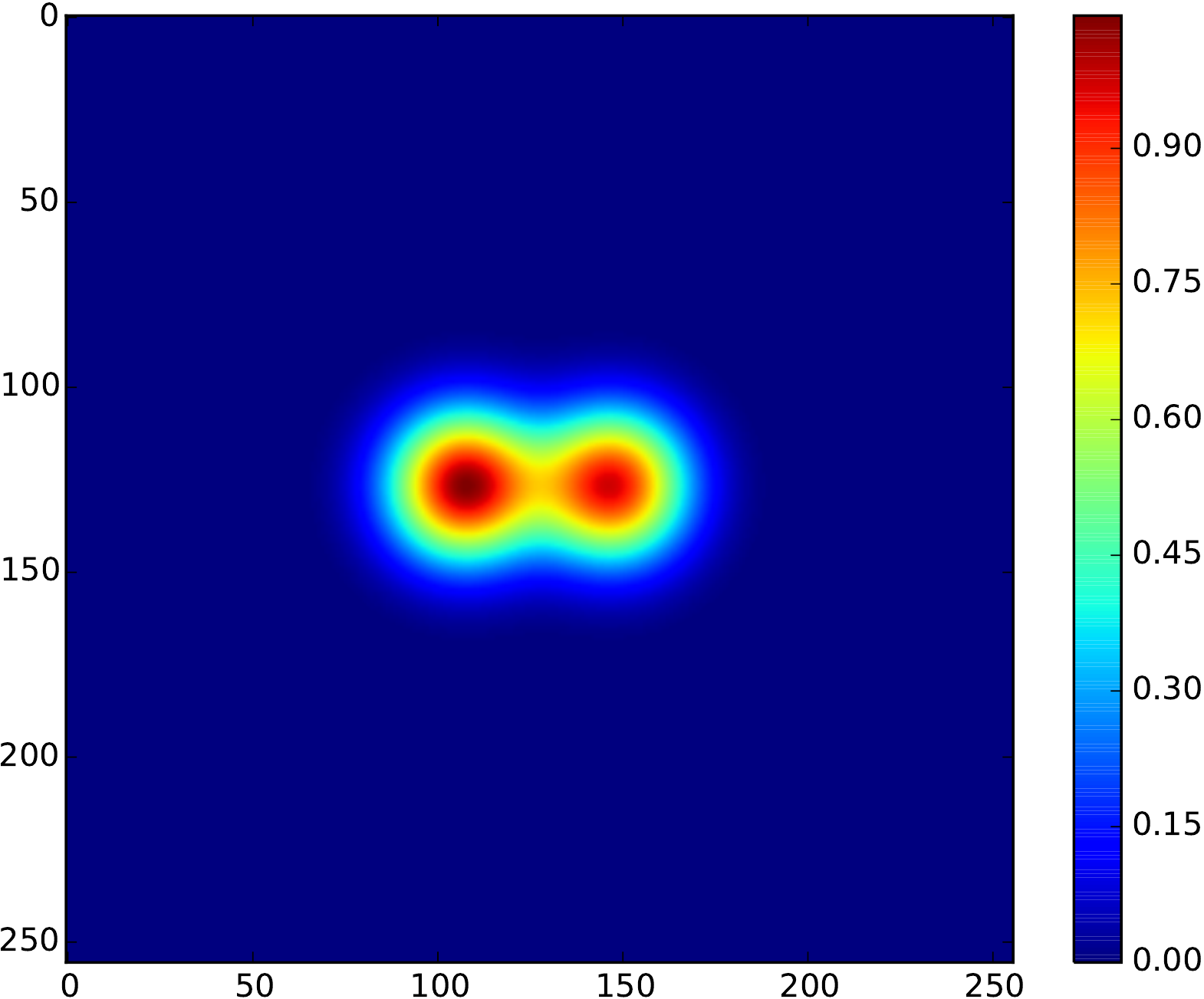}}}
    \caption{ Left. PSF at the central wavelength. Note the large central lobe and high level, ringed sidelobes (up to $40\%$ of the maximum) at large angular distances. 
    			Right. sky object at the central wavelength for simulation 1 : two overlapping Gaussian profiles (in arbitrary flux unit).}
    \label{fig:psf}
\end{figure}

Simulations use PSFs obtained with the HI-inator package\footnote{\texttt{https://github.com/SpheMakh/HI-Inator}} based on MeqTrees software \cite{meqtrees}
with MeerKAT arrays configuration. For the purpose of making Monte Carlo simulations, we simulated small cubes of 15 frequency bands with  images of $256\times 256$ pixels.
 Fig.~\ref{fig:psf} shows the PSF
at the central wavelength, which corresponds to a Fourier coverage produced by a total observation time of 8 hours. In Algorithm 1,
$\bW_s$  in  (\ref{anaylsespat}) corresponds to ``$2^{nd}$'' generation IUWT \cite{Starck07}
and $\bW_s^\dagger$ in (\ref{adjointspat}) is the exact corresponding adjoint operator.

\begin{figure}
	\includegraphics[width= \columnwidth,height = 4.5cm]{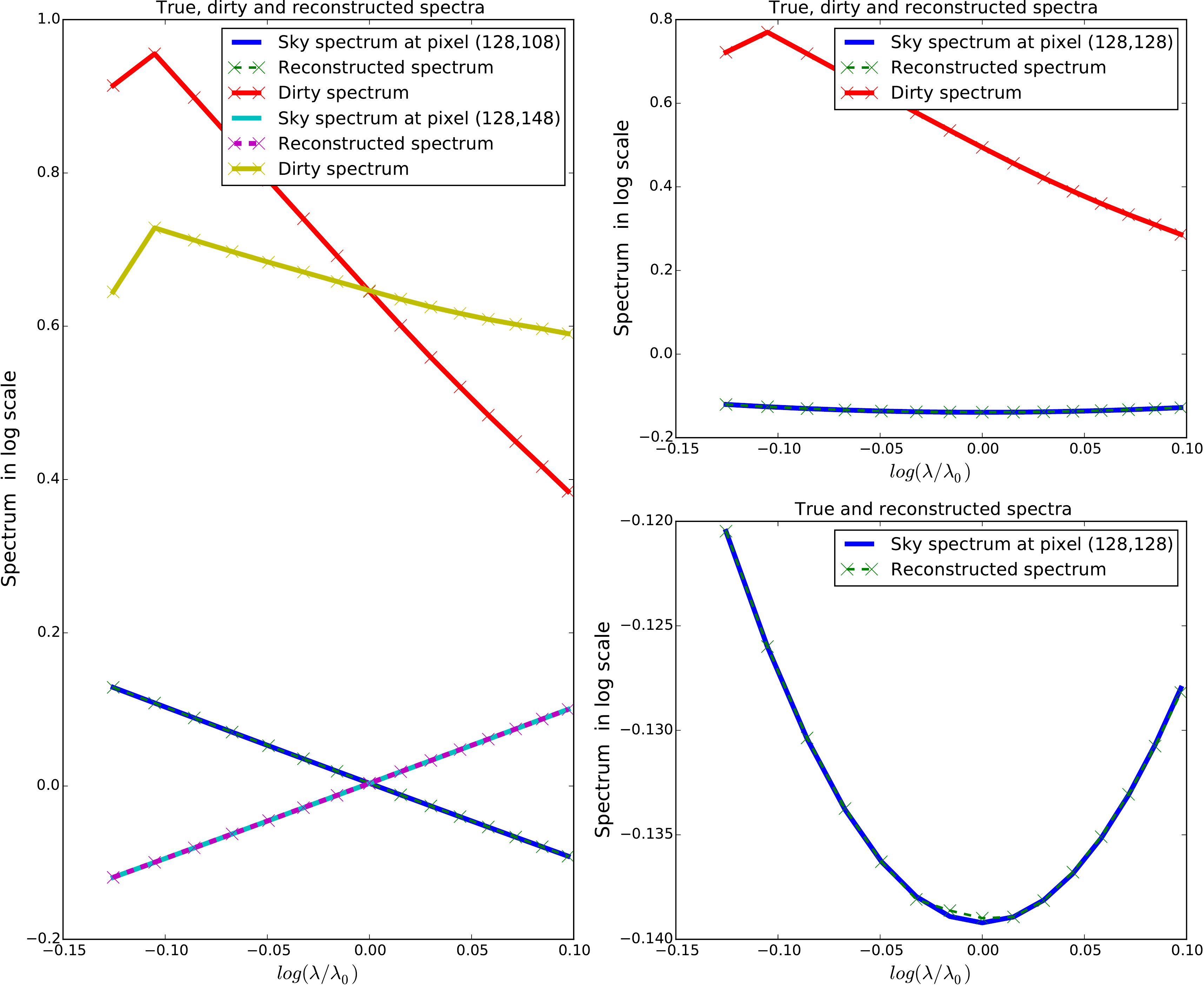}
    \caption{Sky object: two overlapping Gaussian profiles with spectral index equal  to respectively -1.0 and +1.0.
    Left and top right columns represent true, dirty and estimated spectra at different position of the field. Bottom right panel shows the same spectrum as top right but on a different scale. }
    \label{fig:iuwt}
\end{figure}

Two different sky sources are used for these simulations. The first one is similar to the first simulation of \cite{RauMSMF} and is aimed to test  the ability of the algorithm to reconstruct a particular spectrum. 
At a reference wavelength $\lambda_0$ the image consists in two overlapping Gaussian profiles centered at pixel (128,108) and (128,148), see Fig.~\ref{fig:psf} (Right).  The spectra of the two objects are proportional respectively to $\lambda/\lambda_0$ and $\lambda_0/\lambda$ (this corresponds to astronomical 
 spectral indices equal  respectively to $-1.0$ and $+1.0$).  Figure \ref{fig:iuwt} compares the ``dirty'',  true  and  estimated spectra at three spatial positions: pixels
(128,108), (128,128) and  (128,148). 
The left plot shows the results at the two extremal positions. At these positions
 the effect of the most distant object  is negligible: the spectra are proportional to  
 $\lambda/\lambda_0$ and $\lambda_0/\lambda$.  The right plot shows the result obtained 
 at the center of the image: the spectrum is proportional
to  $\lambda/\lambda_0\!\!+\!\!\lambda_0/\lambda$ and cannot be approximated by a simple power law. 
Fig. \ref{fig:iuwt}  shows that this non parametric approach is able  to recover the different types of spectra.

The next simulation is a preliminary result illustrating the relative performances of IUWT w.r.t. the union of eight Daubechies wavelet bases  used in \cite{muffinv1}.
The sky corresponds to the radio emission of an HII region in the M31 galaxy. 
A sky cube is  computed from this real sky image by applying a first order power-law spectrum model. 
 The $256\times 256$ map of spectral indices is constructed following the
procedure detailed in \cite{Junklewitz14}: for each pixel,
the spectral index is a linear combination of an homogeneous Gaussian field and
the reference sky image. A Gaussian noise corresponding to $10$ dB was finally added to the dirty images to simulate instrumental and model errors.
The parameters of the optimization algorithm are set to $\rho  =1$,
$\sigma = 1$ and $\tau = 10^{-5}$. 

A critical problem for the deconvolution of large data cubes is the calibration 
of the regularization parameters $\mu_s$ and $\mu_\lambda$.
We  propose to cope with this problem using the following strategy which decouples the calibration
in two steps:
\begin{enumerate}
\item  $\mu_\lambda$ is first set to $0$: the problem is separable w.r.t. the wavelengths
and each node independently iterates Eqs.~(\ref{grad}-\ref{updatenode}) with $\bt_\ell =
\boldsymbol{0}$. 
This setting which avoids data transfers with the master node  is relatively fast and allows multiple
runs to calibrate $\mu_s$ e.g. by cross-validation.

\item The second step keeps $\mu_s$ and the $\bX$ estimated in step 1) and
calibrates $\mu_\lambda$ using the full algorithm with $\bX$ as an initial condition.

\end{enumerate}

\begin{figure}
	\includegraphics[width=\columnwidth, height=5cm]{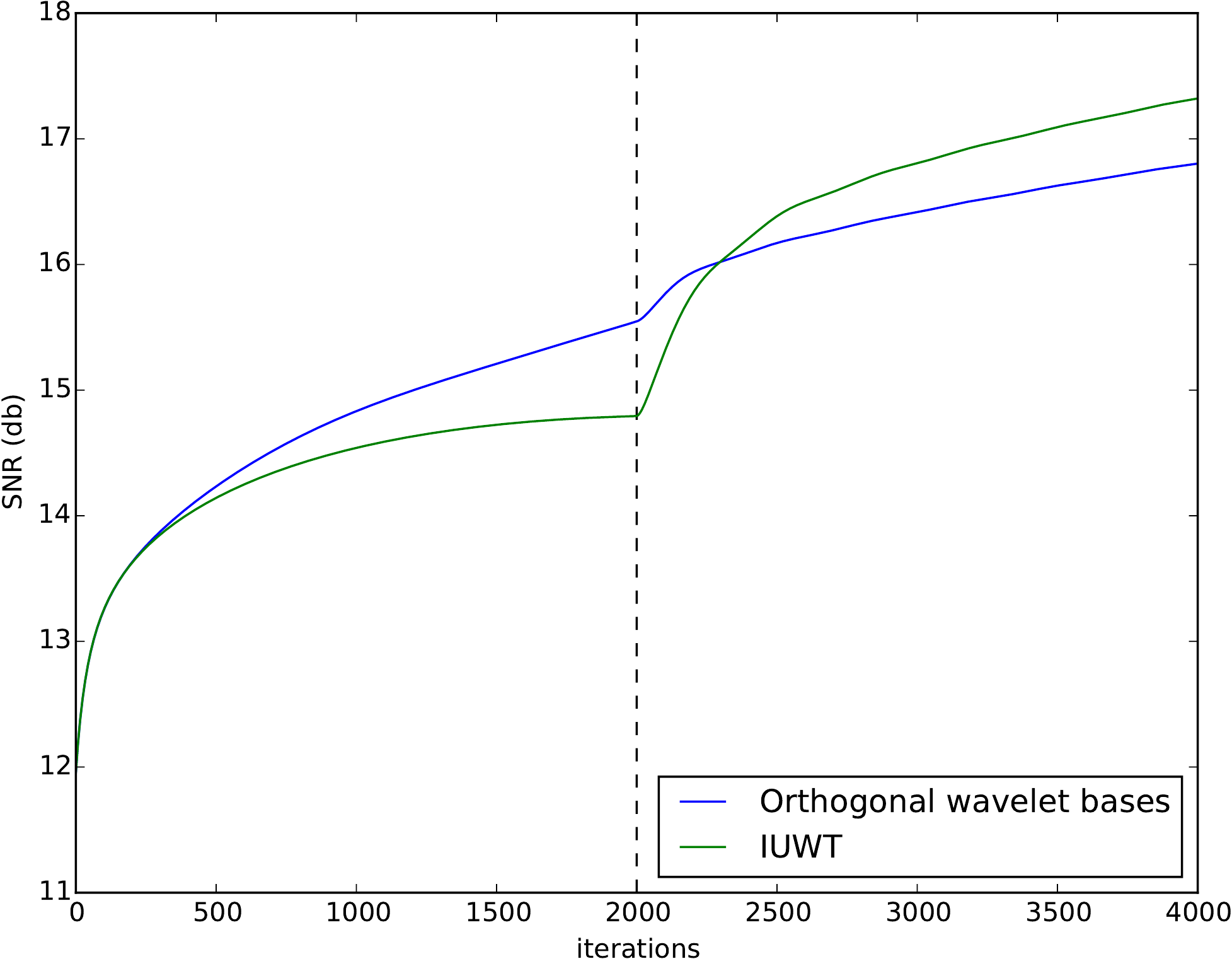}
	\caption{Comparison of the SNR for union of orthogonal bases and IUWT. Spectral regularization
	is turned on at iteration 2000.}
	\label{fig:snr}
\end{figure}
\begin{figure}
\centering
	\includegraphics[width=1.1\columnwidth,angle=90]{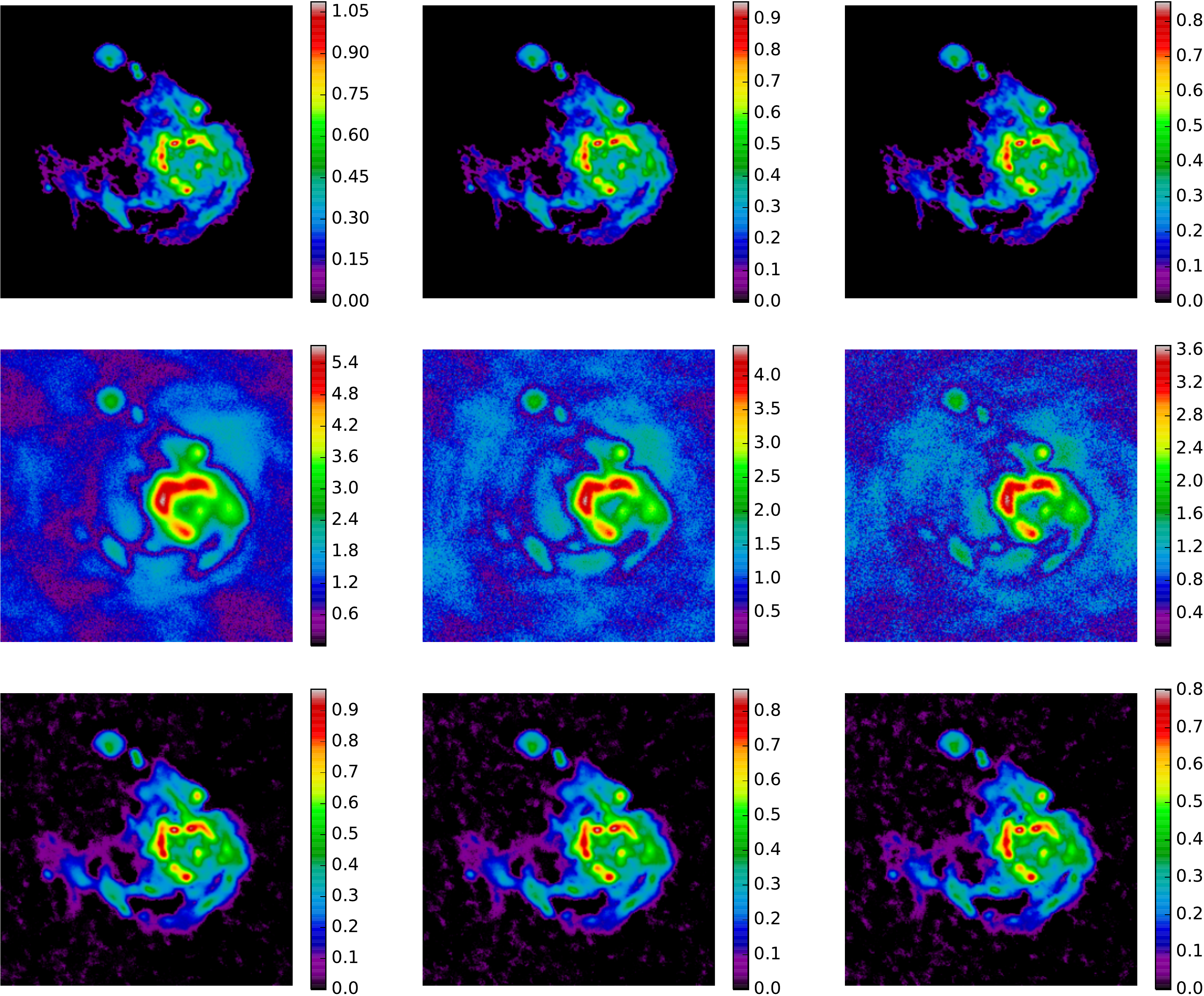}
    \caption{Left column: M31 sky images. Central column: M31 dirty images. Right column: M31 reconstructed images.
    First raw, central raw and bottom raw correspond to the initial, central and last wavelength. }
    \label{fig:recall}
    \vspace{-4mm}
\end{figure}

{
Fig. \ref{fig:snr} compares the  reconstruction Signal to Noise Ratio (SNR) for the union of bases (blue) and IUWT (green) as a function of the iterations.  
SNR is here defined as:}
\begin{equation}
\textrm{SNR}(\bX,\bX^{\bf{\star}}) := 10 \log_{10}\left(\frac{\| {\bX^{\bf{\star}}} \|_{2}^{2}}{\| {\bX} - \bX^{\bf{\star}}  \|_{2}^{2}}\right)
\end{equation}
where ${\bX}$ is the estimated solution and $\bX^{\bf{\star}}$  the ``sky truth''.
The first 2000 iterations correspond to step 1) i.e. $\mu_\lambda = 0$
and $\mu_s = 0.25$, and the following iterations to step 2) i.e. $\mu_s = 0.25$ and 
$\mu_\lambda = 3.0$. The value of $\mu_s=0.25$ in 1) and $\mu_\lambda=3$ in 2) were 
set, for both types of wavelets, after trials and errors in the range $[10^{-4}\; 10^{1}]$ and best performances 
were retained. 

The evolution of the SNRs after iteration 2000, i.e. when $\mu_\lambda>0$ clearly
evidences the gain obtained through a joint spatio-spectral reconstruction for both approaches.
We see that while performances of both approaches are comparable, they relative behavior depend on the regularization and on the number of iterations (which is an important point in a large scale framework). Indeed, such questions deserve further studies. Those are outside the scope of the present paper but are made possible with the parallel implementation proposed in this contribution.

Figure \ref{fig:recall} shows the true sky, the dirty image and the reconstructed image with IUWT at three different wavelengths after 4000 iterations. These results show that the central lobe of the PSF and part of the side lobes, which can be seen in Fig. \ref{fig:psf} (Left),
are properly deconvolved.

Finally, it is worth noting that larger scale tests of  MUFFIN were recently performed (on  data cubes of $2048\times2048\times64$ voxels) 
using a cluster of 8 Xeon E5-26666 compute nodes with 30Gio of memory each. The cluster was built on
AWS using CfnCluster. In this simulation, all variables associated to a single wavelength
on a node use 1GiB of memory and a peak of 4GiB is reached during execution time.

As a conclusion, the proposed alternative implementation of MUFFIN opens the possibility of 
comparing  state-of-the-art sparsity based approaches on large scale spatio-spectral radio imaging problems. 

\bibliographystyle{IEEEtran}

\bibliography{biblio}
\end{document}